\documentstyle[pra,aps,eqsecnum,psfig,epsfig]{revtex}
\begin{document}
\bibliographystyle{unsrt}
\draft

\title{Tunneling of localized excitations: giant enhancement
due to fluctuations }
\author{S. Flach$^1$, V. Fleurov$^2$ and A. A. Ovchinnikov$^1$}
\address{$^1$ MPIPKS, N\"othnitzer Str. 38, 01187 Dresden, Germany \\
$^2$ Beverly and Raymond Sackler Faculty of Exact Sciences\\
School of Physics and Astronomy, Tel Aviv University\\
Tel Aviv 69978, Israel}

\date{\today}
\maketitle

\begin{abstract}
We consider the tunneling of localized excitations (many boson bound states)
in the presence of a bosonic bath. We show both analytically and numerically
that the bath influence results in a dramatical enhancement of the amplitude
of the excitation tunneling. The order of the bosonic flow in the course of
the tunneling process is obtained. On the background of the giant tunneling
enhancement we observe and describe additional resonant enhancement and
{\sl suppression} of tunneling due to avoided level crossings.
\end{abstract}

\baselineskip 19pt

\section{Introduction}

Tunneling properties of, e.g., local bond excitations in molecules, of bound
states in quantum lattices or of large spins in the presence of anisotropy
fields share a lot of similarities of corresponding model approaches and
technical treatments. Despite the amount of published work this field lacks
complete understanding. Especially intriguing are the relation to so-called
dynamical tunneling in phase space \cite{btu93}, the influence of heat bath
coupling \cite{lcdfgz87,w99}, the exact computation of tunneling probabilities
\cite{mw86,wh87}, to name a few.

To study different aspects of the problem it is helpful to formulate a model
which shares most of the needed properties, and is numerically solvable
(because it is typically hard to obtain exact analytical results).

In the present paper we will study the tunneling of multi boson bound states
in a model which we believe posesses all necessary qualitative ingredients
needed for the understanding of a more general case.

The Hamiltonian of the model is given by
\begin{equation}
H=\frac{1}{2} \left[ \left(a_1^+a_1\right)^2 +
 \left(a_2^+a_2\right)^2 \right] +
 C\left( a_1^+a_2 + a_1 a_2^+\right) + \delta
\left(  a_1^+a_3 + a_1 a_3^+ +a_2^+a_3 + a_2 a_3^+ \right)
\label{1-1}
\end{equation}
Here the operators $a_l,a_l^+$ ($l=1,2,3$) are bosonic annihilation and
creation operators with standard commutation relations
$[a_l,a_m^+]=\delta_{lm}$. Note that (\ref{1-1}) is invariant under
permutation of site 1 with site 2.

Equation (\ref{1-1}) describes, e.g., a three site (or bond) molecule, i.e. a
trimer \cite{sfvf97}. It could also serve as a model for tunneling in a dimer
\cite{afko96} (just sites 1 and 2) under the simultaneous influence of a heat
(or energy) bath on site 3, coupled to the symmetric modes of the dimer. It
may be used for a study of excitations in $H_2O$ molecules, where the two
$O-H$ stretching modes (dimer) are additionally coupled to the relative angle
between the two bonds (third site) \cite{oe82}.

Finally (\ref{1-1}) may be transformed into a model of a spin with local
anisotropy and external field (site 3) coupled to a bosonic field (using
Schwinger bosons for sites 1 and 2). The coupling is such that the total spin
length will not be conserved, but may fluctuate. The dimer limit $\delta =0$
may be represented as a spin length conserving model with the Hamiltonian $H_s
= S_z^2 + 2CS_x$ \cite{dag91,gcs00}.

Model (\ref{1-1}) conserves in addition to the energy also the total particle
number $b$ which is an eigenvalue of the total number operator $B$
\begin{equation}
B= n_1 +n_2 +n_3\;\;,\;\;n_l = a_l^+ a_l \;\;.
\label{1-2}
\end{equation}
Note that due to this conservation the infinite dimensional Hilbert space
decomposes into an infinite set of finite dimensional orthogonal subspaces,
each with a given total particle number. Consequently numerical evaluations
will always translate into diagonalizations of finite matrices, without
worrying about truncation errors. The classical limit is obtained by replacing
the operators $a_l$ with complex scalars $\Psi_l$ (Hermitian conjugation is
replaced by complex conjugation).

Besides the above mentioned general interest in such kinds of model we want to
give some additional reasons for studing of (\ref{1-1}). During the past
decade we witnessed an explosion of interest to a new class of excitations in
lattice models. These models are typically described by Hamiltonian equations
of motion of degrees of freedom associated with lattice points of spatial
periodic lattices. In the presence of interaction between these degrees of
freedom, when considering linear equations of motion, the translational
symmetry of the underlying lattice allows one to obtain delocalized plane wave
solutions (phonons, magnons and other {\sl whateverons}. The spectra of these
excitations have finite upper bounds due to the discreteness of the system. If
instead considering rather arbitrary additional nonlinearities in the
equations of motion, a new generic type of solutions appears: {\bf discrete
breathers} \cite{sfcrw98}. These discrete breather solutions are time-periodic
spatially localized excitations and appear in one-parameter families. The
discrete breather concept can be easily extended to dissipative systems
\cite{sm97}. 

Much more complicated is the quantization of these excitations. The question
arises which eigenstates of the quantum Hamiltonian operator correspond to
classical discrete breathers. Intuition leads to the conclusion that one
should search for multi {\sl whateveron} bound states \cite{rsm00}. One
signature should be the appearance of extremely narrow quasiparticle bands in
the high energy sector. 'Quasiparticle' stands for the need of just one wave
vector which labels the states of the quantum breather band (as opposed to
many wave vectors which label many {\sl whateveron} continua)
\cite{wgbs96,wbgs98}. This view is supported by the consideration of a
(semiclassical) tunneling process of a discrete breather, which implicitely
assumes that the quantum system may support coherent tunneling of localized
excitations similar to a (quasi)particle \cite{fsf98}. 'Extremely narrow'
could have different meanings. In the best case these bands should be much
narrower than the mean level spacing in the given energy sector. This seems to
be not so simple owing to the increase of the density of states when going up
in energy. Another aspect of 'narrow' means weak interaction with nearby lying
states.

In contrast to the classical case, where implicit and explicit existence
proofs of discrete breathers are known
\cite{ma94,sf95-pre-1,lsm97,cbrsm97}, and where reliable computational
tools are available wich allows one to obtain numerical solutions on large
lattices\cite{sfcrw98}, the situation with the quantum case is
different. Except for some integrable one-dimensional models no analytical
solutions are known. Numerical evaluations will have to face the tremendous
problem of diagonalizing large matrices even for moderate system sizes (like
6-10 sites). In order to gain an insight into the problem it is useful to
study small systems where the discrete breather concept can be partially
used. Although there is no localization length anymore, we may still find
excitations which are mainly located on one site. The smallest possible system
is thus a two site system, i.e. a dimer. For $\delta =0$ equation (\ref{1-1})
describes an integrable dimer. The analysis of this model \cite{afko96}
demonstrates that the existence of classical discrete breather solutions
corresponds to the appearance of tunneling pairs of eigenstates. These pairs
of eigenstates have energy splittings which are orders of magnitude smaller
than the averaged level spacing. The relative splittings decrease with
the increase of the eigenenergies, in full correspondence with the above
described expectations.

One nongeneric feature of the dimer model is its integrability. Although it
allowed to obtain analytical results, it could hide some features of the
generic nonintegrable case. This conclusion led to the consideration of the
trimer model. The third site was deliberately chosen to be different from the
first two sites in order to preserve the symmetries of the model. Some
classical and quantum properties of (\ref{1-1}) were described in detail in a
previous paper \cite{sfvf97}. The discussion of the classical properties was
devoted mainly to periodic orbit bifurcations which allow for an appearance of
discrete breather solutions transforming into each other under the
$1\leftrightarrow 2$ permutation in the trimer. In the quantum case we again
observe pairs of symmetric and antisymmetric states with nearly coinciding
energy levels. Note that the energy splitting in such pairs is determined by
the phase space tunneling amplitude between the above two classical solutions.

The paper \cite{sfvf97} presents data on the behavior of such pairs as well
as of individual levels with varying coupling parameter $\delta $. We observe
that pairs can be crossed by other pairs and/or by individual levels. The
crossing for levels with the same symmetry is certainly avoided whereas levels
with differing symmetries really cross each other. In many cases a pair is
able to survive many crossings and restore itself up to large values of
$\delta $. However, sooner or later a fatal crossing comes which destroys the
pair. In the classical system this loss is manifested by a strong overlap of
the corresponding quantum wave function with chaotic layers in the phase space
of the classical model.

The level splittings were not analized in more detail. Instead only their
behaviour on scales comparable with the averaged level spacing were of
interest. Here we will present a careful analysis of the mighty changes of the
tiny splittings as the coupling $\delta$ is changed. Denote with $x,y,z$ the
eigenvalues of the site number operators $n_1,n_2,n_3$. We may consider the
quantum states of the trimer at $\delta =0$ when $z$ is a good quantum number
and then follow the evolution of these states with increasing $\delta $. The
state for $\delta =0$ can be traced back to $C=0$ and be thus characterized in
addition by $x$ and $y$.

\section{Tunneling order, magnitude and resonances}

In Fig. 2(b) in the previous paper \cite{sfvf97} we picked a tunneling pair
which has energy $\approx 342$ for $\delta =0$. The fixed model parameters
were $B=40$ and $C=2$. The pair states are characterized by $x=26 (0)$, $y=0
(26)$ and $z=14$ for $C=\delta =0$. Note that this pair survives approximately
30 avoided crossings before it finally is destroyed at coupling strength
$\delta \approx 2.67$.

Figure 1 presents the dependence of the tunneling splitting, $\Delta E$, for
this pair. One sees that the splitting rapidly increases gaining about eight
orders of magnitude when $\delta$ changes from 0 to slightly above 0.5. Then
this rapid but nevertheless smooth rise is interrupted by very sharp spikes
when the splitting $\Delta E$ rises by several orders of magnitude with
$\delta $ changing by mere percents and then abruptly changes in the opposite
direction sometimes even overshooting its prespike value. Such spikes, some
larger, some smaller, repeat with increasing $\delta $ untill the splitting
value approaches the mean level spacing (of order one in the figure). Only
then one may say that the pair is destroyed since it can be hardly
distinguished among the other trimer levels.
\begin{figure}[htb]
\psfig{file=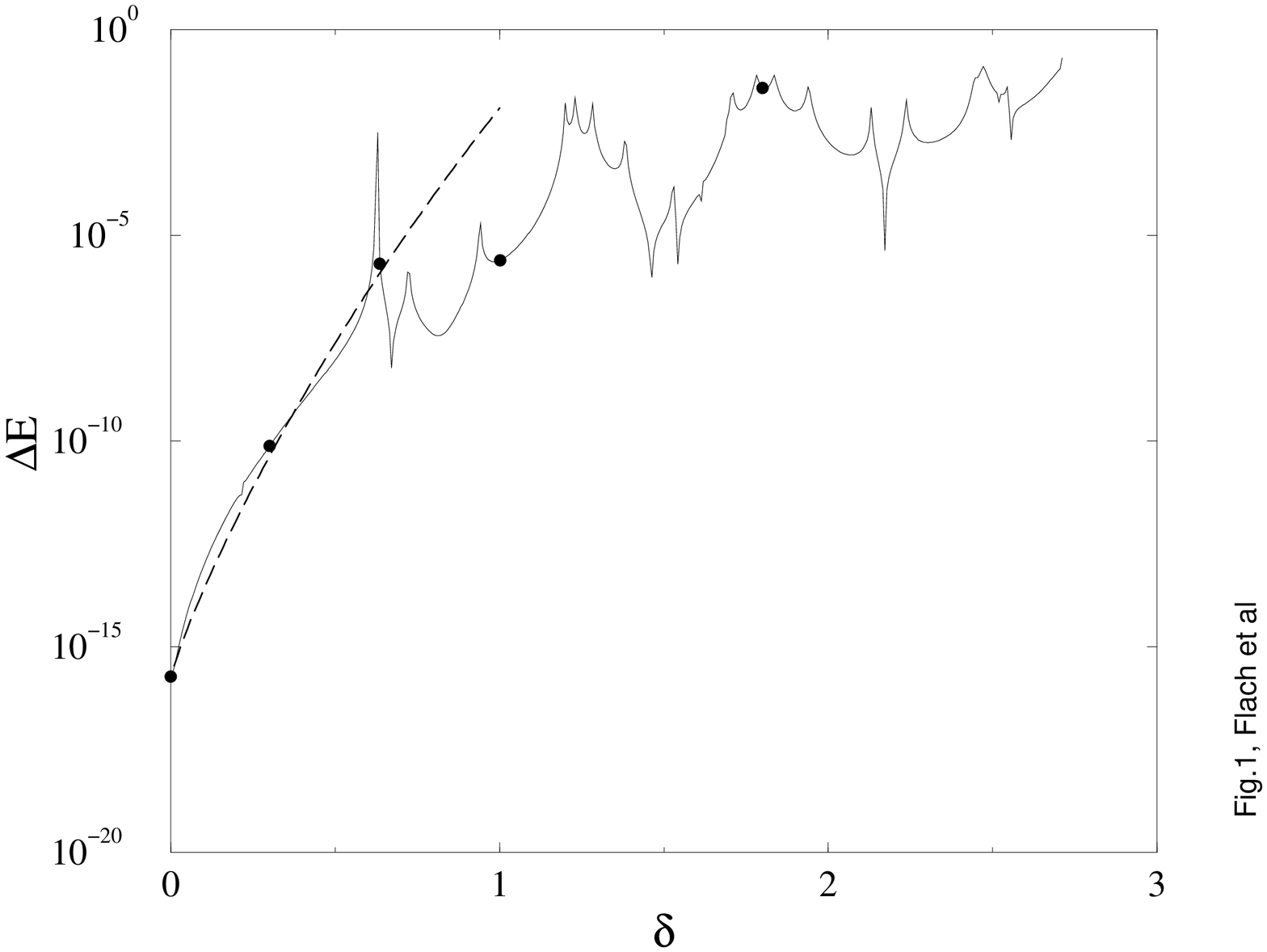,width=3.4in,angle=270}
\caption{
Level splitting versus $\delta$ for level pair as described in the text.
Solid line - numerical result.
Dashed line - semiclassical approximation as described in text.
Filled circles - location of wave function analysis in Fig.2.}
\label{fig1}
\end{figure}

Another observation is presented in Figs.2 (a-e). We plot the intensity
distribution of the logarithm of the squared symmetric wavefunction of our
chosen pair for five different values of $\delta=0\;,\;0.3\;,\;0.636\;,\;
1.0\;,\;1.8$ (their locations are indicated by filled circles in Fig.1). We
use the eigenstates of $B$ as basis states. They can be formally represented
as $|x,y,z>$ where $x,y,z$ are the particle numbers on sites 1, 2, 3,
respectively. Due to the commutation of $B$ with $H$ two site occupation
numbers are enough if the total particle number is fixed. Thus the final
encoding of states (for a given value of $b$) can be chosen as $|x,z)$ (see
also \cite{sfvf97} for details). The abscissa in Figs.2 is $x$ and the
ordinate is $z$. Thus the intensity plots provide us with information about
the order of particle flow in the course the tunneling process.

\begin{figure}[htb]
\psfig{file=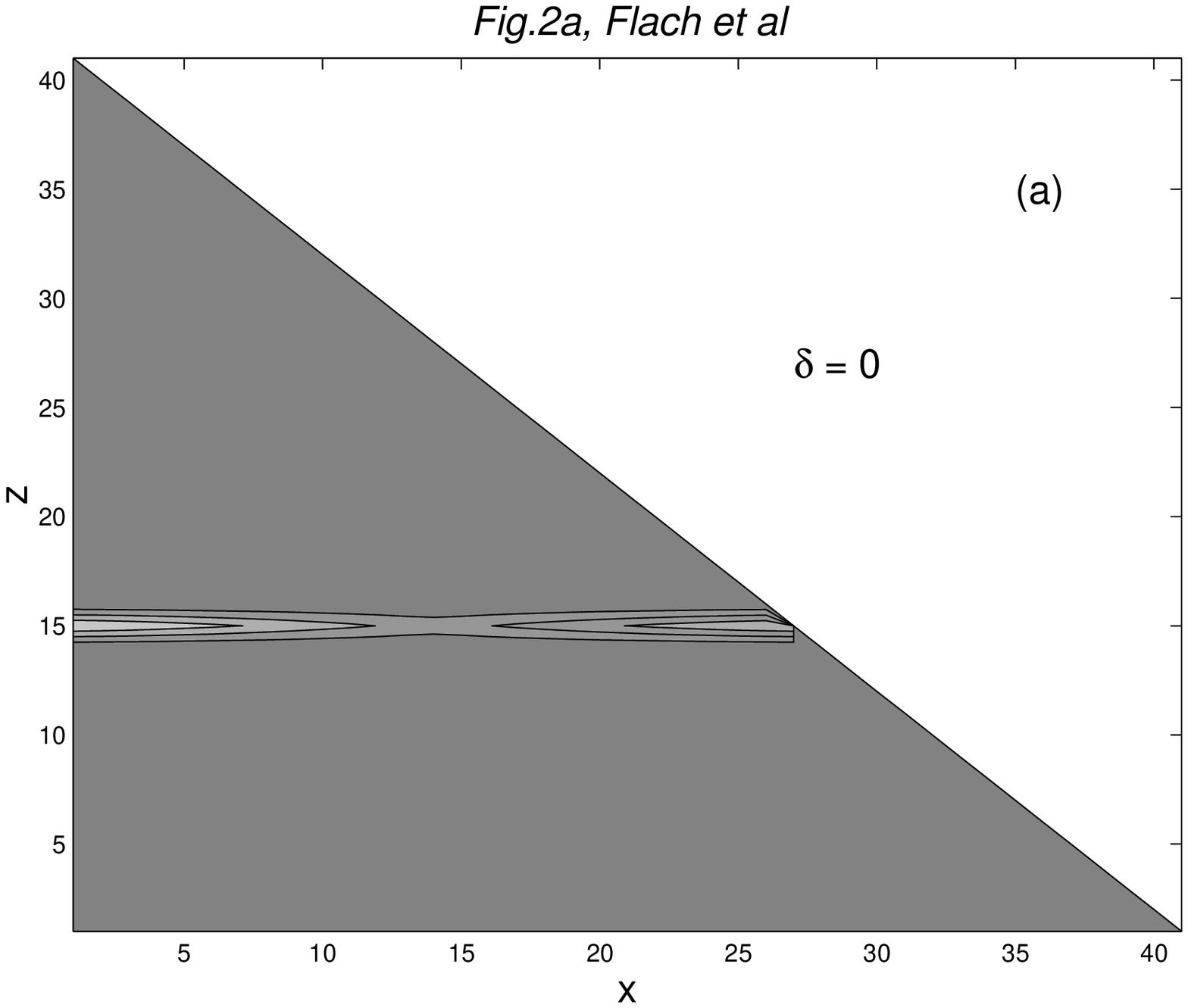,width=3.4in,angle=0}
\label{fig2a}
\end{figure}
\begin{figure}[htb]
\vspace*{-8.2cm}
\hspace*{9cm}
\psfig{file=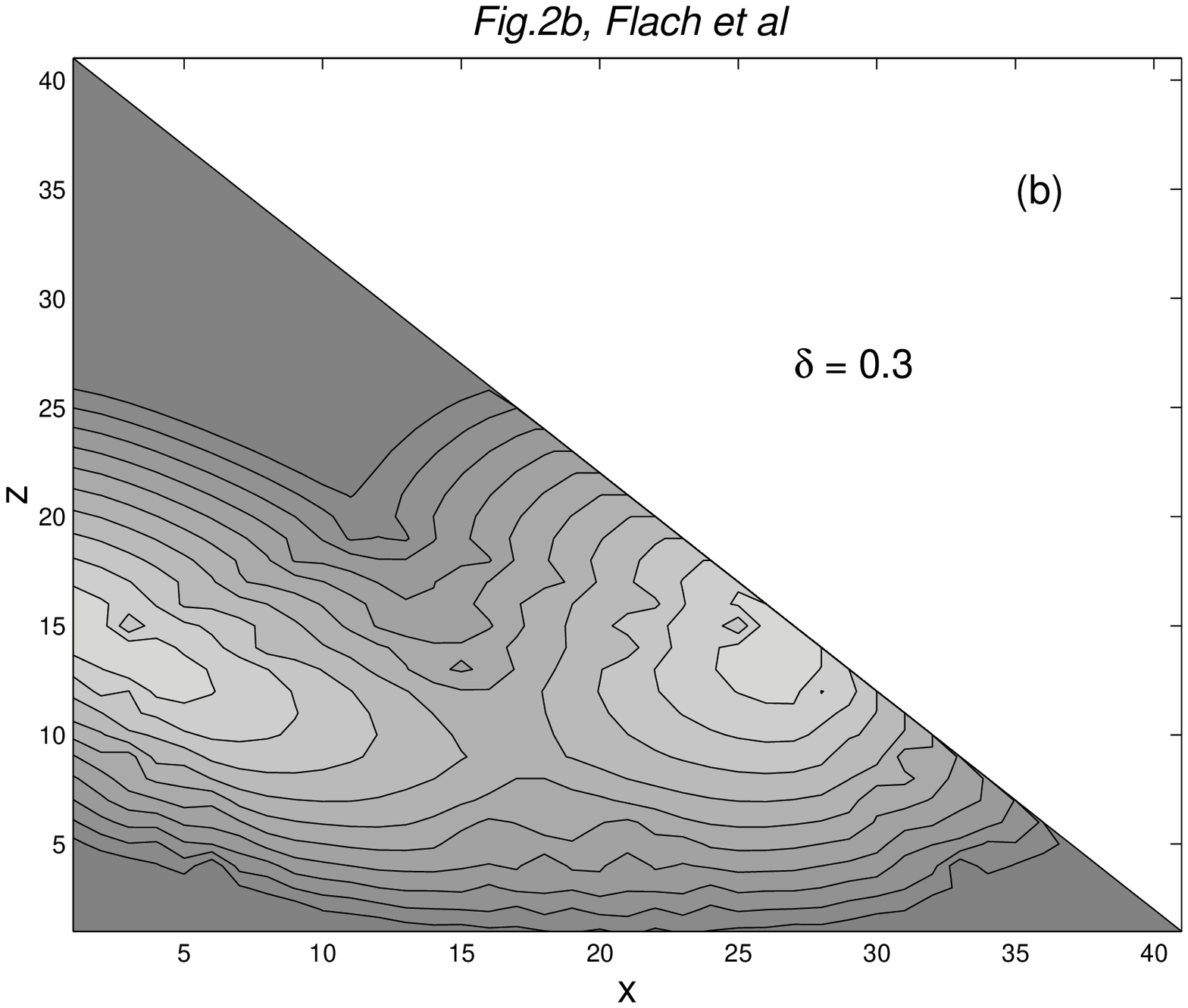,width=3.4in,angle=0}
\label{fig2b}
\end{figure}

\begin{figure}[htb]
\psfig{file=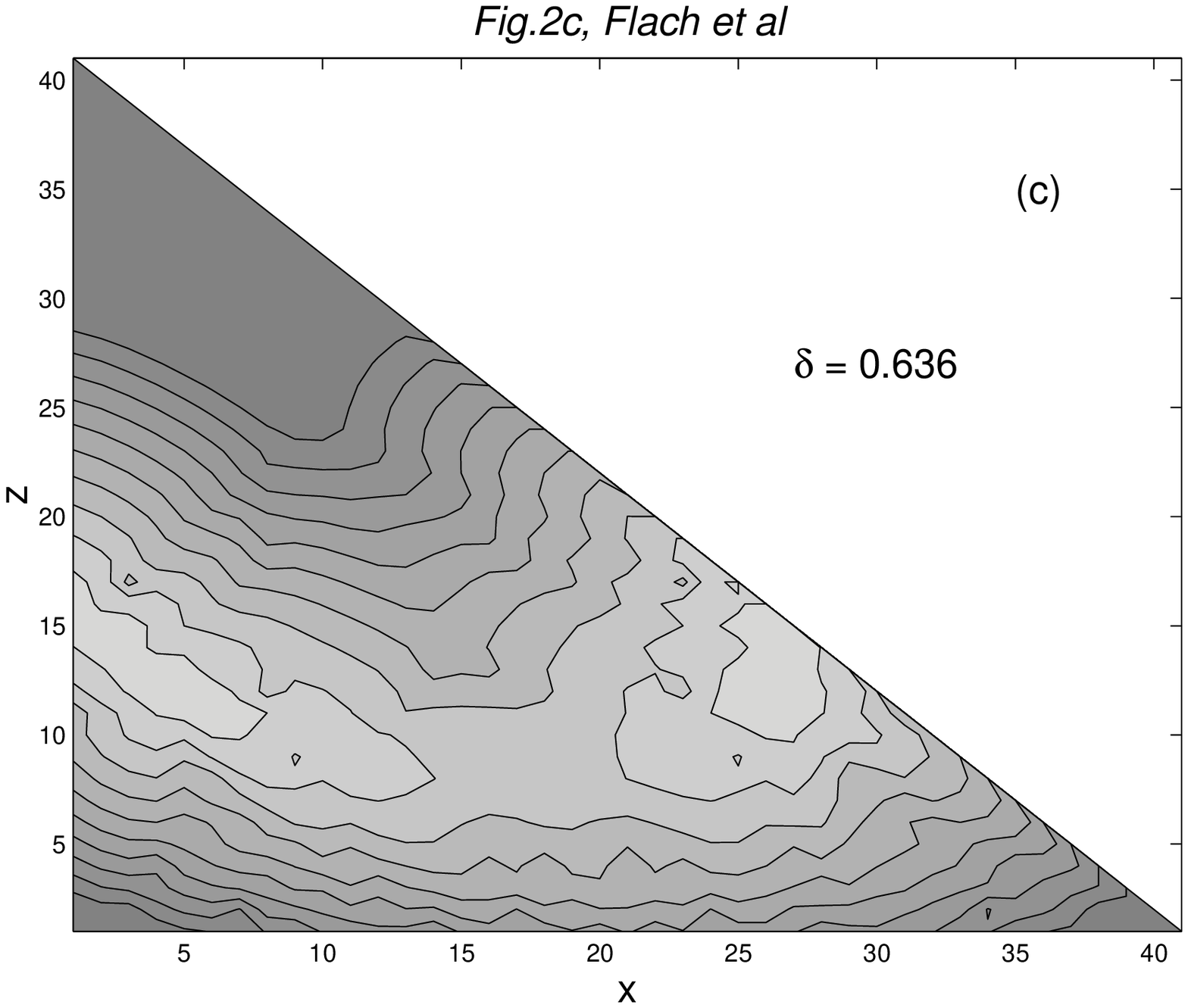,width=3.4in,angle=0}
\label{fig2c}
\end{figure}
\begin{figure}[htb]
\vspace*{-8.2cm}
\hspace*{9cm}
\psfig{file=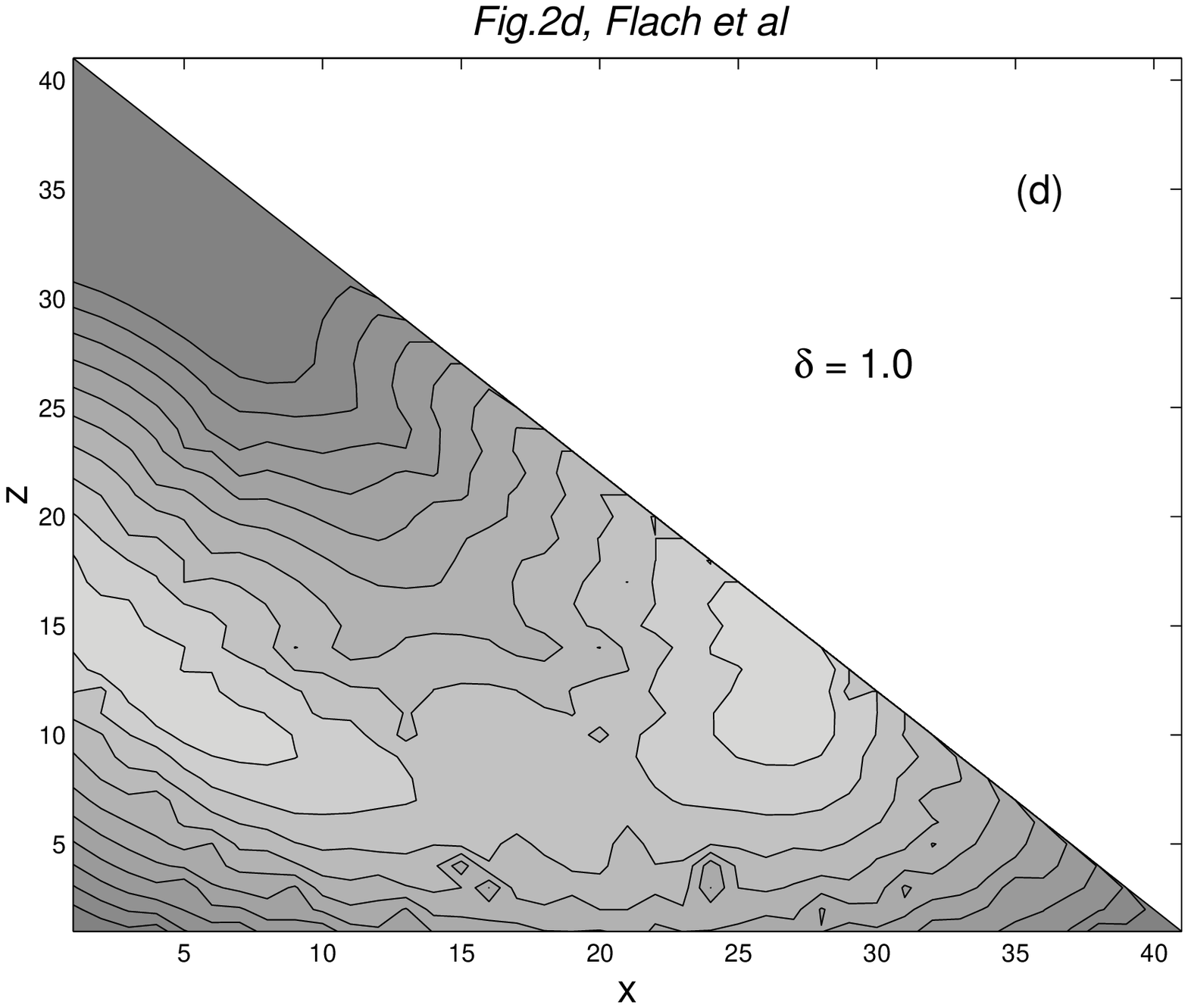,width=3.4in,angle=0}
\label{fig2d}
\end{figure}
\begin{figure}[htb]
\psfig{file=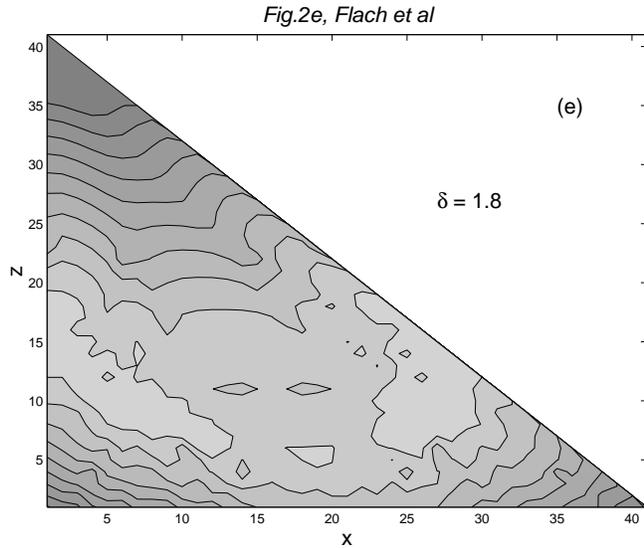,width=3.4in,angle=0}
\caption{
Countour plot of the logarithm of the symmetric eigenstate of the chosen
tunneling pair (cf. Fig.1) for five different values of
$\delta=0,\;0.3,\;0.636,\;1.0,\;1.8$ (their location is indicated by
filled circles in Fig.1). 
\\
In Fig.2(a) three equidistant grid lines are used. In Figs.2(b-e) ten grid
lines are used. Minimum value of squared wave function is $10^{-30}$, maximum
value is about 1. 
}
\label{fig2e}
\end{figure}

For $\delta =0$ (Fig.2a) the only possibility for the 26 particles on site 1
is to directly tunnel to site 2. Site 3 is decoupled with its 14 particles not
participating in the process. The squared wavefunction takes the form of a
compact rim in the $(x,z)$ plane which is parallel to the $x$ axis. Nonzero
values of the wavefunction are observed only on the rim. This direct tunneling
has been quantitatively described in \cite{afko96}. When switching on some
nonzero coupling to the third site, the particle number on the dimer (sites
1,2) is not conserved anymore. The third site serves as a particle reservoir
which is able either to collect particles from or supply particles to the
dimer. This coupling will allow for nonzero values of the wavefunction away
from the rim. But most importantly, it will change the shape of the rim. We
observe that the rim is bended down to smaller $z$ values with increasing
$\delta$. That implies that the order of tunneling (when, e.g., going from
large to small $x$ values) is as follows: first, some particles tunnel from
site 1 to site 2 and simultaneously from site 3 to site 2
(Fig.3(a)). Afterwards particles flow from site 1 to both sites 2 and 3
(Fig.3(b)). With increasing $\delta$ the structure of the wavefunction
intensity becomes more and more complex, possibly revealing information about
the classical phase space flow structure.
\begin{figure}[htb]
\psfig{file=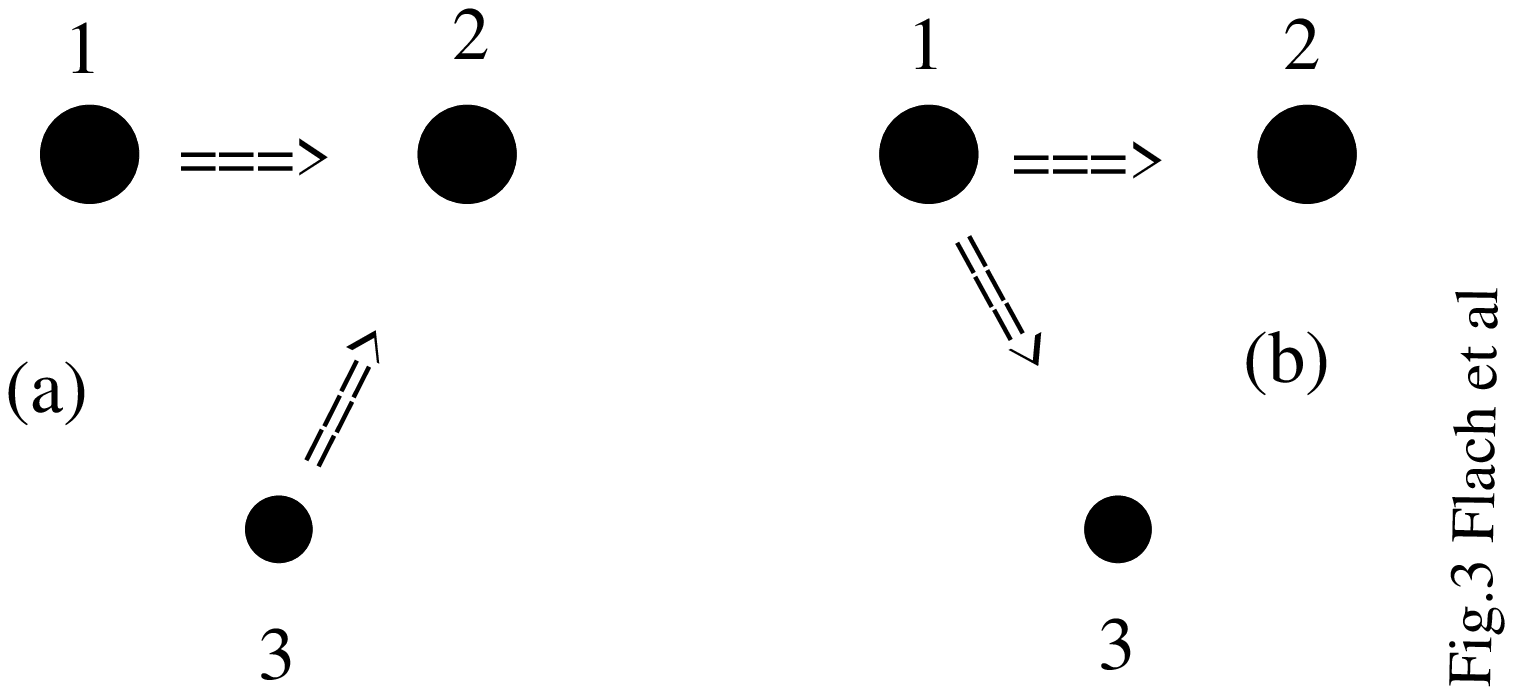,width=2.5in,angle=270}
\caption{
Order of tunneling in the trimer. Filled large circles - sites 1 and 2, filled
small circle - site 3. Arrows indicate direction of transfer of particles.
}
\label{fig3}
\end{figure}

Thus we observe three intriguing features. First, the tunneling splitting
increases by eight orders of magnitude when $\delta$ increases from zero to
0.5. This seems to be unexpected, since at those values perturbation theory in
$\delta$ should be applicable (at least Fig.2b in \cite{sfvf97} demonstrates
that this should be true for the levels themselfes). The second observation is
that the tunneling begins with a flow of particles from the bath (site 3)
directly to the empty site which is to be filled (with simultaneous flow from
the filled dimer site to the empty one). At the end of the tunneling process
the initially filled dimer site is giving particles back to the bath
site. Again this is an unexpected result, since it implies that the particle
number on the dimer is increasing during the tunneling, which seems to
decrease the tunneling probability, according to the results for an isolated
dimer. These first two results are closely  connected, as we will show
below. The third result concernes the resonant structure on top of the smooth
variation in Fig.1. The resonant enhancements and suppressions of tunneling
are related to avoided crossings. Their presence implies that a fine tuning of
the system parameters may strongly suppress or enhance tunneling which may be
useful for spectroscopic devices. In the following we will explain all three
observations.

\subsection{The order of tunneling}

To understand the order of tunneling we consider a generalized model
Hamiltonian for $C=\delta=0$
\begin{equation}
H_g = f(n_1) + f(n_2) + g(n_3) \;\;.
\label{2-1}
\end{equation}
It reduces to (\ref{1-1}) when choosing $f(x)=x^2/2$ and $g(z)=0$. Switching
on some interaction inside the dimer ($C \neq 0$) we will observe some
tunneling probability. The smallness of this probability is due to the
presence of an energy barrier hindering successive transferring of particles
from one site of the dimer to the other. This energy barrier is given by
\begin{equation}
\Delta _b = \max_{0 < x < (b-z_0)} |f(b-z_0) - (f(b-z_0-x) + f(x)) |
\label{2-2}
\end{equation}
where the total number of particles $b$ and the number of particles on the
still isolated third site $z_0$ are fixed. Note that the top of the barrier is
reached when the particles are evenly spread inside the dimer (it is
straightforward to generalize to other cases).

At nonzero $\delta$ the value of $z$ may fluctuate. Now we have to consider
the energy in the two-dimensional plane $(x,z)$. The energy landscape for an
arbitrary distribution of particles with respect to the chosen initial
distribution $x=0$, $y=b-z_0$, $z=z_0$ reads
\begin{equation}
\Delta _b (x,z) = | f(x) + f(b-x-z) + g(z) - f(b-z_0) - g(z_0) | \;\;.
\label{2-3}
\end{equation}
For small values of $\delta$ the variation of $z$ will be small and we may
expand around $z_0$: $z=z_0 + \delta z$. As noted above (\ref{2-3}) has a
maximum at $x=(b-z_0)/2$ for fixed $z_0$. In general the derivative of this
function along $z$ will, however, not vanish in this point. This implies that
a variation of $z$ will change the barrier height. We have to determine the
sign of the variation $\delta z$ which lowers the barrier height. This sign
will tell us whether the particle number on site 3 will decrease or increase
during the tunneling. Thus, we may predict or explain the change of the rim
shape of the above described squared wavefunction. This in turn will allow us
to conclude about the order of tunneling. Define $e_1=f(b-z_0)-2f((b-z_0)/2)$
and $e_2=f'((b-z_0)/2) - g'(z_0)$. Straightforward arithmetics leads to the
result:
\begin{eqnarray}
\delta z > 0& \;\; {\rm if} \;\;[e_1 >0\;{\rm and}\; e_2 < 0]\;{\rm or}\;
[e_1<0\;{\rm and}\;e_2 > 0]
\label{2-4} \\
\delta z < 0 & \;\;{\rm otherwise}
\end{eqnarray}
Our chosen Hamiltonian (\ref{1-1}) yields $\delta z < 0$ which is in full
accord with the observed change of the rim shape in Figs. 2.

The above considerations allow us to conclude that the tunneling enhancement
due to the lowering of the barrier is not sensitive to the sign of $\delta
z$. The only possibility to avoid tunneling enhancement is to choose
$e_2=0$. Still one will be faced with consideration of higher order
contributions of $\delta z$ to the barrier change.

\subsection{Quantitative account for the tunneling enhancement}

Now we present a semiclassical approach which is capable of accounting for the
rapid increase of the energy splitting in the tunneling pair shown in
Fig.1. For this sake we shall first discuss tunneling in the dimer. Then we
will show how the introduction of an interaction with the third site will
modify the results.

A semiclassical treatment of tunneling in the dimer was outlined in
\cite{fsf98} where canonical coordinates were introduced, using the classical
variables $\Psi_l = \sqrt{n_l}{\rm e}^{i \varphi_l}$,
\begin{eqnarray}
q=\frac{1}{2}[n_2 - n_1], & p=\varphi_2 - \varphi_1,  \nonumber \\
&  \nonumber \\
J = n_1 + n_2, &\ \ \ \chi= \frac{1}{2} \left(\varphi_1 +
\varphi_2\right)\;\;.  \label{3-1} 
\end{eqnarray}

These coordinates allow one to represent the dimer Hamiltonian in the form
\begin{equation}
H_d=\left({\frac{J}{2}}\right)^2+q^2+
2C\sqrt{\left({\frac{J}{2}}\right)^2-q^2 }\cos p  \label{3-2}
\end{equation}
Here the phase $p$ plays the role of a momentum whereas $q$ is the coordinate
along which the tunneling in the dimer takes place. The coordinate $J$ is an
integral of motion, since the Hamiltonian does not depend on its canonically
conjugated momentum $\chi$. In order to calculate the tunneling probability
amplitude we first choose two symmetry breaking isolated periodic orbits with
a certain energy $E$ and then look for a classical path $q(\tau)$ in the
complex time plane which connects these two orbits. On this way we have to
cross classically forbidden areas where the momentum $p$ becomes
imaginary. What we need to know is the dependence of the momentum $p$ on the
coordinate $q$ which, in the classically forbidden area, is determined by the
equation
\begin{equation}
E=\left({\frac{J}{2}}\right)^2+q_0^2+
2C\sqrt{\left({\frac{J}{2}}\right)^2-q_0^2}\cosh{p}.
\label{3-3}
\end{equation}
The boundaries $\pm q_0(E)$ of this area or the turning points are obtained
from equation (\ref{3-3}) at $p=0$ for a given energy $E$.

The tunneling probability amplitude can be estimated as 
\[
W\sim \exp \{-S\} 
\]
where 
\begin{equation}
S=
\displaystyle \int
\limits_{-q_{0}}^{q_{0}}p (q)dq.  \label{3-4}
\end{equation}
This integral can be readily evaluated for the energy $E_{3}=J^{2}/2+C^{2}$
(for a given value of $J$). Then $q_{0}=\sqrt{J^{2}/4-C^{2}}$ and
\begin{equation}
S(E=E_{3})=-\sqrt{J^{2}-4C^{2}}+
J\ln \left( \frac{2C}{J-\sqrt{J^{2}-4C^{2}}} \right)  \label{3-5}
\end{equation}

This expression coincides, to within the preexponential factor, with the
results obtained in \cite{fsf98}. However our aim here is to consider the
tunneling amplitude of the trimer or, to be more specific, consider how the
tunneling splitting for a certain pair of the trimer levels varies with the
increase of the coupling parameter $\delta$.

A naive approach is based on equation (\ref{3-5}). An interaction with the
third site of the trimer may lead to a change of the number $J$ of particles
in the dimer (first two sites). One can readily see that the action
(\ref{3-5}) becomes larger for larger $J$ values. Since the total number of
particles in the trimer is fixed, $b=J+n_3$, an increase of $J$ occurs only at
the expense of the third site which may provide the necessary
particles. Therefore the tunneling amplitude decreases if $n_3$ should
decrease, and this is completely opposite to what we have observed numerically
in the previous section. It means that a more detailed analysis is needed.

First, we introduce canonical coordinates which are explicitly symmetric with
respect to permutation of the first two sites of the trimer.
\begin{eqnarray}
Q= \frac{1}{3} (n_1 + n_2) - \frac{2}{3} n_3, &\ \ \ \pi = \frac{1}{2}
(\varphi_1 + \varphi_2) - \varphi_3,  \nonumber \\
&  \nonumber \\
b = n_1 + n_2 + n_3,\ \ \ \ \ \ & \bar \chi = \frac{1}{3} (\varphi_1 +
\varphi_2 + \varphi_3).  \label{3-6}
\end{eqnarray}
The definitions of the two coordinates $q$ and $p$, related to the dimer,
remain as in equation (\ref{3-1}). Then the trimer Hamiltonian (\ref{1-1})
takes the form
\begin{eqnarray}
H= & \left(\frac{1}{3}b + \frac{1}{2} Q\right)^2 + q^2 +
2C\sqrt{\left(\frac{1}{3}b + \frac{1}{2} Q\right)^2 - q^2} \cos p  \nonumber
\\
&  \nonumber \\
& + 2\delta \sqrt{\left(\frac{1}{3}b - Q\right) \left(\frac{1}{3} b +
\frac{1}{2} Q + q \right)} \cos\left(\pi + \frac{1}{2} p \right)
\nonumber \\
&  \nonumber \\
& + 2\delta \sqrt{\left(\frac{1}{3}b - Q\right) \left(\frac{1}{3} b +
\frac{1}{2} z - q \right)} \cos\left(\pi - \frac{1}{2} p \right).
\label{3-7}
\end{eqnarray}
It becomes straightforward to obtain equations of motions for the four
relevant coordinates, with $b$ being an integral of motion and $\bar\chi$
redundant. We need here only two of this equations
\begin{eqnarray}
\dot Q = & - 2\delta \sqrt{\left(\frac{1}{3}b - Q\right) \left(\frac{1}{3} b +
\frac{1}{2} Q + q \right)} \sin\left(\pi + \frac{1}{2} p \right)
\nonumber \\
&  \nonumber \\
& - 2\delta \sqrt{\left(\frac{1}{3}b - Q\right) \left(\frac{1}{3} b +
\frac{1}{2} Q - q \right)} \sin \left(\pi - \frac{1}{2} p \right)
\label{3-8}
\end{eqnarray}
\begin{eqnarray}
-\dot\pi=& \frac{1}{3}b + \frac{1}{2} Q +
\displaystyle{{\frac{C\left(\frac{1}{3}b +
\frac{1}{2} Q \right)}{\sqrt{\left(\frac{1}{3}b + \frac{1}{2}
Q\right)^2 - q^2}}}} \cos p+  \nonumber \\
&  \nonumber \\
& \delta \cos \left(\pi + \frac{1}{2} p \right) \left[\frac{1}{2}
\sqrt{\displaystyle{\frac{\left(\frac{1}{3}b - Q\right)}{\left(\frac{1}{3} b +
\frac{1}{2} Q + q \right)}}} - \sqrt{\displaystyle{\frac{\left(\frac{1}{3} b
+\frac{1}{2} Q + q \right) }{\left(\frac{1}{3}b - q\right)}}}\right]+
\nonumber \\
&  \nonumber \\
& \delta \cos \left(\pi - \frac{1}{2} p \right) \left[\frac{1}{2}
\sqrt{\displaystyle{\frac{\left(\frac{1}{3}b - Q\right)}{\left(\frac{1}{3} b +
\frac{1}{2} Q - q \right)}}} - \sqrt{\displaystyle{\frac{\left(\frac{1}{3} b +\frac{1}{2} Q - q \right) }{\left(\frac{1}{3}b - Q\right)}}}\right].
\label{3-9}
\end{eqnarray}

These equations will allow us to estimate the deviation of the tunneling
trajectory in the ($q,\ Q$) plane. The latter can be directly mapped onto the
($z$, $x$) plane. Again we may consider trajectories connecting isolated
periodic orbits with $\dot p=p=\dot\pi=\pi=0$. A reasonable approximation is
to assume that $\dot\pi=\pi=0$ along the path and then use equation
(\ref{3-8}) to estimate the deviation of the $Q$ value from its initial $Q_0$
values,
\begin{equation}
\Delta Q \sim \dot Q_{max}\frac{\Delta\tau}{2}\sim -
\frac{\delta\sqrt{Q_0(b/3-Q_0)}}{\sqrt{2}C}  \label{3-10}
\end{equation}
Here $\dot Q_{max}$ is the largest value of $\dot Q$. It is estimated by
substituting $Q=Q_0$, $q=0$ (the middle point of the tunneling path) and
$p_{max}$ into equation (\ref{3-8}). As for $p_{max}$ it is found for the
given dimer energy from equation (\ref{3-2}) at $q=0$. $\Delta\tau =
1/\sqrt{J^2/4-C^2}$ is the time of motion through the forbidden region
(traversal time).

First, we again obtain that the tunneling trajectory really deviates towards
lower values of $Q$, i.e. towards a decreasing number $z$ of bosons in the
third site. It fits our numerical results. In order to calculate the
corresponding action we slightly deform the trajectory. Better to say we
straighten the trajectory, which finally simply becomes a straight line with
constant $z$ which is shifted towards lower $z$-values. We believe that this
deformation would lead to a negligible change of the result since the
deformation is carried out mainly in the borders of the forbidden area. The
principal contribution to the integral comes, however, from the central part
of this area.

This procedure allows us to return to the calculation of the tunneling
amplitude in the dimer by means of the integral (\ref{3-4}). However, the
number of bosons, which has been $J$ in the isolated dimer, becomes now
$J^{\prime }=J+\Delta Q$. As for the energy of the dimer it should remain
without change ($E=E_{3}$), since we neglect small corrections (order of
$\delta ^{2}$) due to the interaction with third site. The distance to the
upper bound $E_{3}^{\prime }=J^{\prime}{}^{2}/2+C^{4}$ can be readily found,
$\Delta E\approx J\Delta Q$. There should be also a small shift of the turning
points $q_{0}^{\prime }$ which is, however, of a minor importance.

As a result we can represent the action in an integral form
\begin{equation}
S= \displaystyle \int\limits_{-q_{0}^{\prime }}^{q_{0}^{\prime }}dq\ \arccos
{\frac{{{{\displaystyle {J^{\prime }{}^{2} \over 4}}} + \Delta
E-q^{2}}}{{2C\sqrt{{{\displaystyle {J^{\prime }{}^{2} \over 4}}}-q^{2}}}}}
\label{3-11}
\end{equation}
This integral can be calculated analytically and in the trivial case of
$\Delta Q=0$ it reduces to equation (\ref{3-5}). However, the general analytic
expression for (ref{3-11}) is extremely cumbersome. That is why we have
preferred to calculate this integral numerically. The thick dashed line in
Fig.1 is the graphical form of (\ref{3-11}). We observe very good agreement
with the numerical data for tunneling in the trimer up to $\delta$ values
where possibly our expansion becomes too crude and where the spiky structure
sets in. Yet our consideration is capable of quantitatively describing the
raising of the tunneling splitting by eight orders of magnitude!

\subsection{Resonances due to avoided crossings}

Let us give a semiquantitative understanding of the spiky behaviour of the
energy splitting dependence on $\delta$ in Fig.1. Comparing the position of
the spikes with the level variation in Fig.2b in \cite{sfvf97} we immediately
realize that each resonance in Fig.1 is related to an avoided crossing of our
level pair with either a single third level or another tunneling pair. A
simple analytical description of such crossing can be obtained following the
same simple reasoning which has led to the understanding of an avoided
crossing of just two levels (see, e.g., \cite{LLIII}).

Let us first assume that we have found all eigenenergies and eigenfunctions of
the trimer for given values of the parameters $C$ and $\delta _{0}$. We denote
the energies and wave function of our tunneling pair as $\varepsilon_{1s}$,
$\varepsilon _{1a}$ and $\psi _{1s}$, $\psi _{1a}$. Here indices $s$ and $a$
relate to the symmetric and antisymmetric states. We also assume that there is
a level or another pair of levels which happen to be very close to our
tunneling pair. Then we may consider small variations $\tilde{\delta}$ of the
coupling parameter $\delta $ near its $\delta _{0}$ value ($\delta =\delta_{0}
+ \tilde{\delta}$) and calculate the variation of the level energies by means
of the perturbation theory, neglecting contributions from all other more
distant levels. The corresponding perturbation is explicitly symmetric with
respect to the $1-2$ permutations. Therefore it may mix only the states with
the same symmetry, so that those levels will avoid crossings. As for states of
different symmetries, their crossing is allowed.

\subsubsection{Pair --- single level crossing}

First we consider the case when the tunneling pair is approached by a single
level which we choose to be symmetric with respect to the permutation of the
dimer sites. The energy and the wave function of this level at $\delta =
\delta _{0}$ are $\varepsilon _{2s}$ and $\psi _{2s}$. Such a crossing can be
described by the $3\times 3$ matrix
\begin{equation}
\left(
\begin{array}{ccc}
\varepsilon _{1s}+V_{s11}\tilde{\delta} & V_{s12}\tilde{\delta} & 0 \\ 
&  &  \\ 
V_{s12}\tilde{\delta} & \varepsilon _{2s}+V_{s22} & 0 \\ 
&  &  \\ 
0 & 0 & \varepsilon _{1a}+V_{a11}
\end{array}
\right) .  \label{3-13}
\end{equation}
where all terms proportional to $\tilde{\delta}$ are the corresponding matrix
elements of the perturbation. They vanish at $\delta_0$ which is close enough
to the avoided crossing.

Diagonalizing the matrix (\ref{3-13}) we obtain  three levels
\begin{eqnarray}
E_{s\{1,2\}} &=&{\frac{\varepsilon _{s1}+\varepsilon _{s2}+(V_{s11}+V_{s22})
\tilde{\delta}}{2}}\pm \sqrt{{\frac{[\varepsilon _{s1}-\varepsilon
_{s2}+(V_{s11}-V_{s22})\tilde{\delta}]^{2}}{4}}+
V_{s12}^{2}\tilde{\delta}^{2}}  \nonumber \\
&&  \nonumber \\
E_{a} &=&\varepsilon _{a1}+V_{a11}\tilde{\delta}\hspace{10cm}  \label{3-14}
\end{eqnarray}

There are two different scenaria which depend on the parameters of the
matrix. In the first case at $\delta=\delta_0$ the order of the levels in
either increasing or decreasing order is $\varepsilon_{1a}$,
$\varepsilon_{1s}$, $\varepsilon_{2s}$. In this case we will first see a
decrease of the splitting down to zero (!) and a subsequent increase of the
splitting up to some maximum with subsequent decrease down to its original
value. In the second case $\varepsilon_{2s}$, $\varepsilon{1a}$,
$\varepsilon{1s}$ the above sequence is reversed - we first observe an
increase up to a maximum, then a decrease down to zero and a final increase to
the original splitting value. All the parameters may be chosen to be
positive. The difference between the coefficients $V_{s11}$ and $V_{a11}$ is
neglected since this difference reflects the slow change of the level
splitting with varying $\delta$ (cf. Fig.1). The perturbation theory allows us
to deal only with small values of $\tilde{\delta} $, hence parameters should
satisfy the conditions that $|\varepsilon_{s1}-\varepsilon _{s2}|\ll
|V_{s11}-V_{s22}|$ and $|V_{s12}|\ll 1$. 

Then the tunneling splitting $\Delta $ can be found as a function of
$\tilde{\delta}$. It can be defined as the smallest distance between the antisymmetric level and the two symmetric levels, 
\begin{equation}
\Delta =\min \{|E_{s1}-E_{a}|,|E_{s2}-E_{a}|\}  \label{3-15}
\end{equation}

\begin{figure}[htb]
\psfig{file=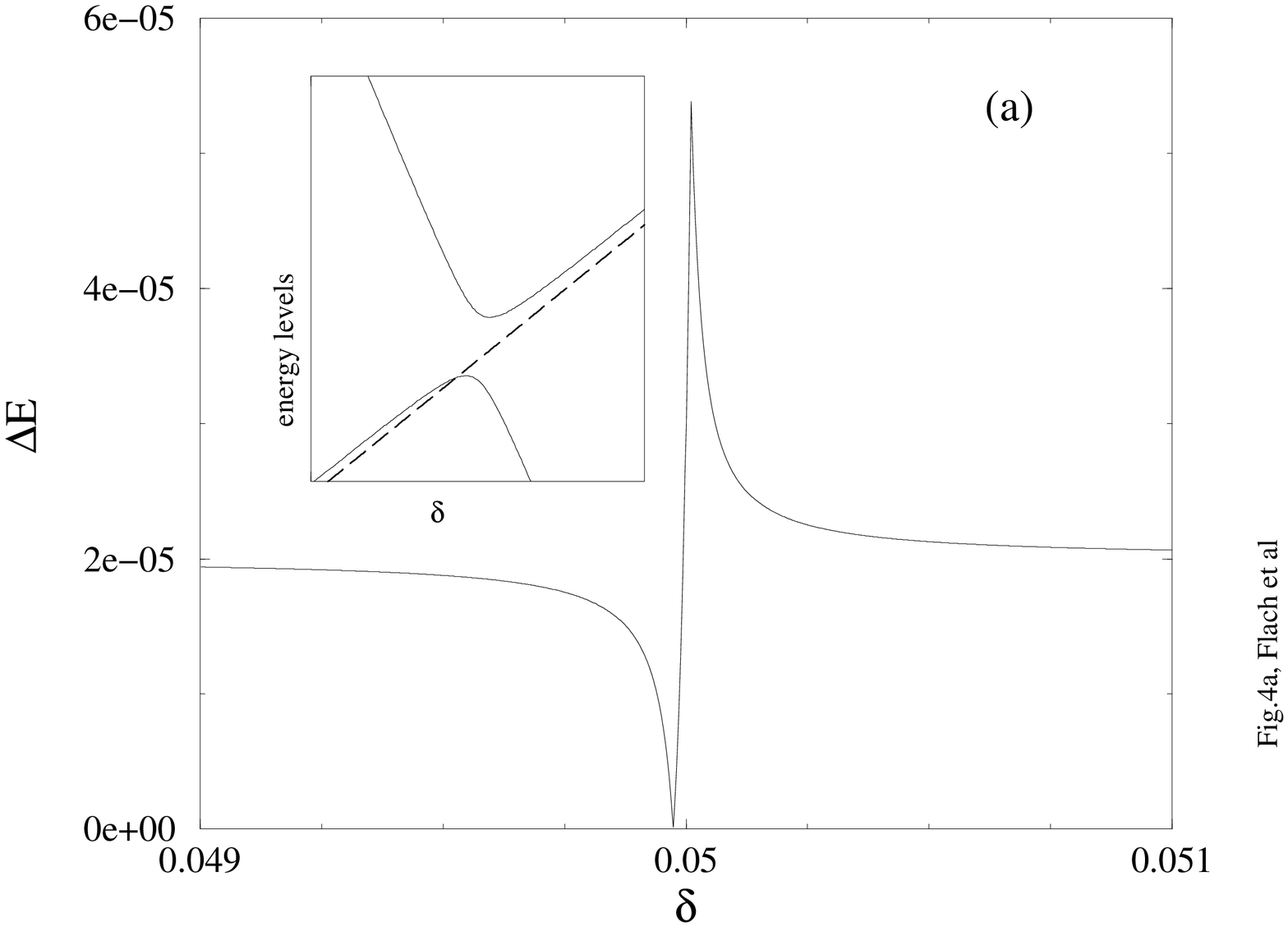,width=2.5in,angle=270}
\label{fig4a}
\end{figure}
\begin{figure}[htb]
\vspace*{-7.2cm}
\hspace*{9cm}
\psfig{file=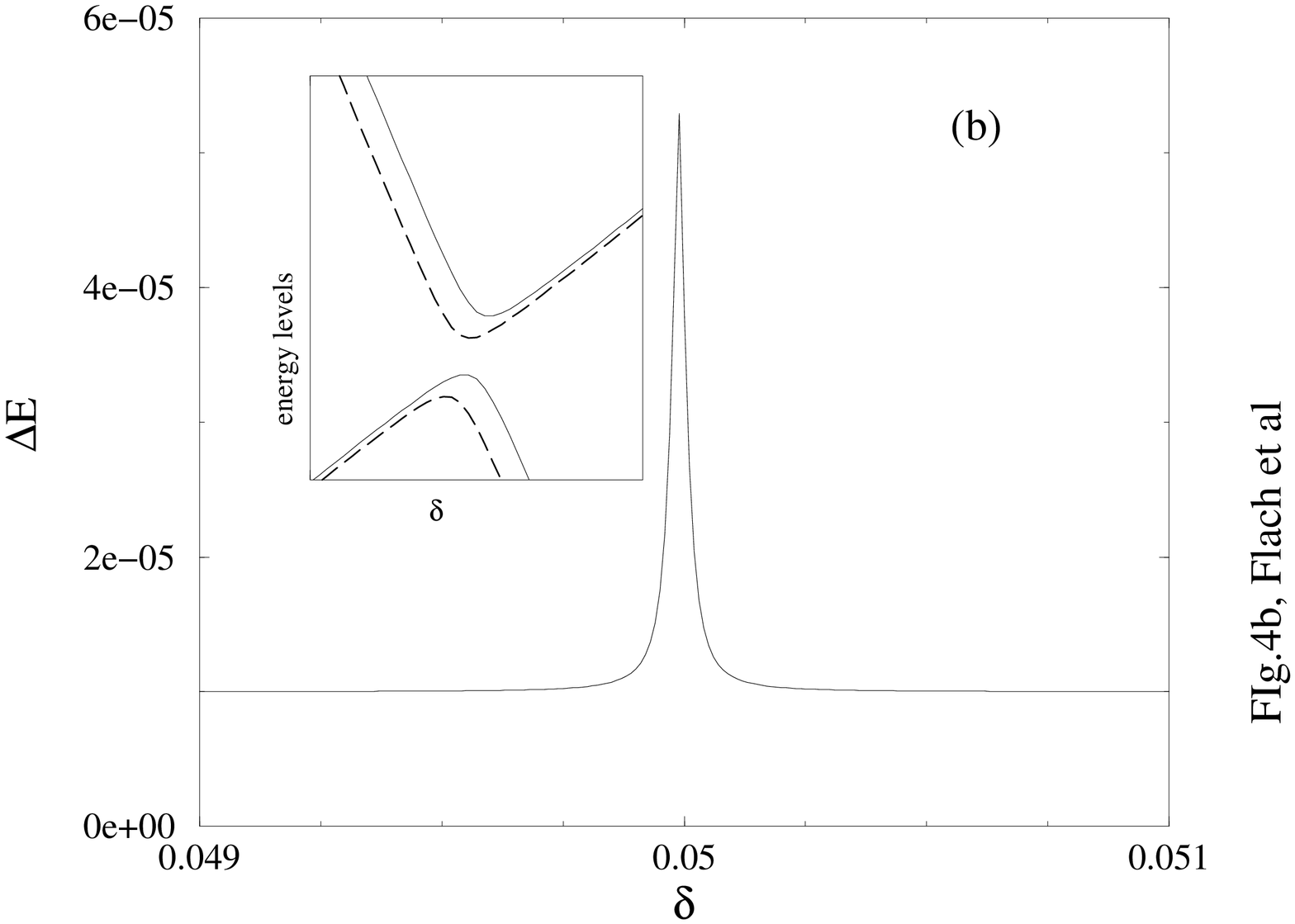,width=2.5in,angle=270}
\label{fig4b}
\end{figure}
\begin{figure}[htb]
\psfig{file=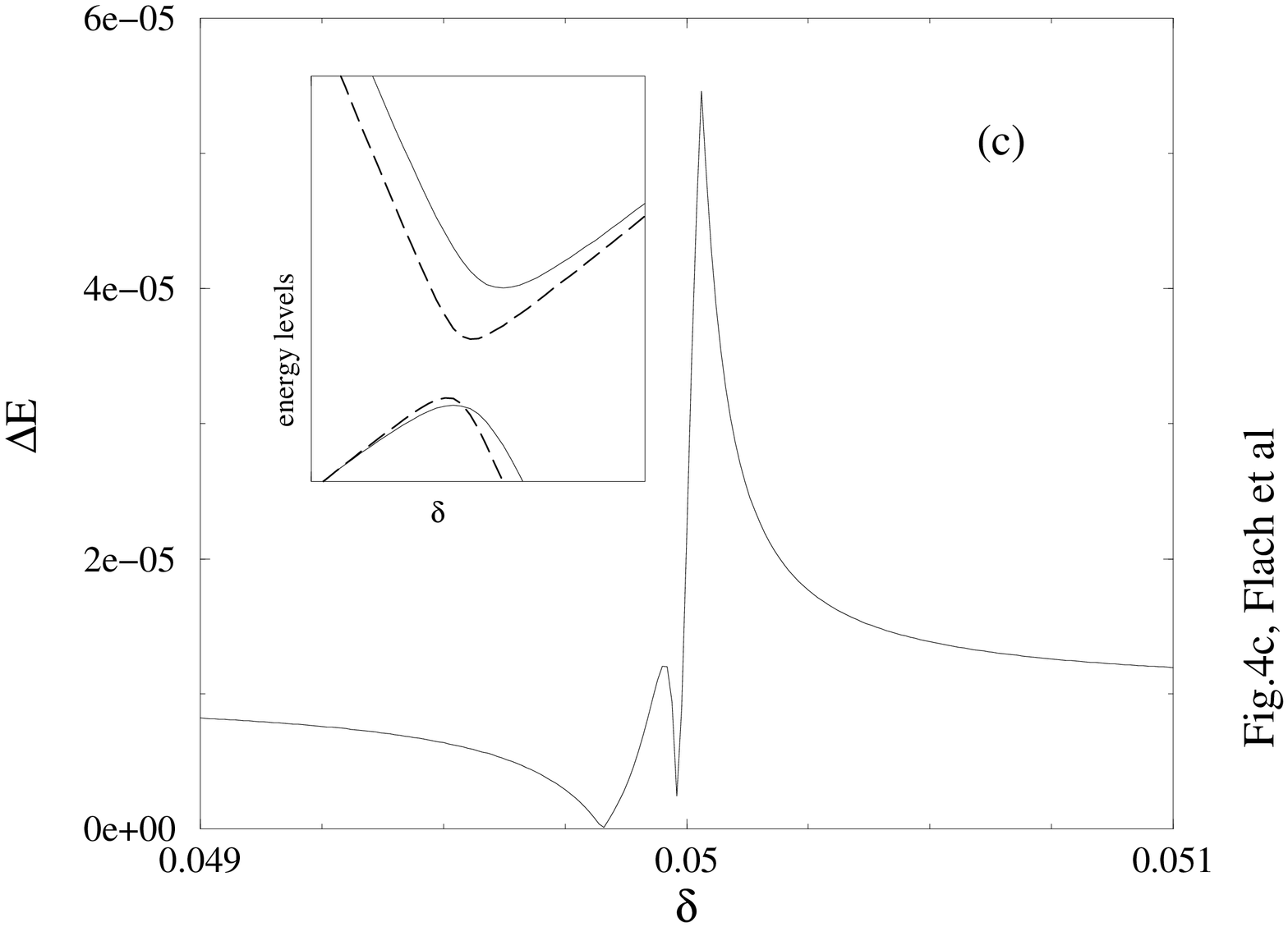,width=2.5in,angle=270}
\label{fig4c}
\end{figure}
\begin{figure}[htb]
\vspace*{-7.2cm}
\hspace*{9cm}
\psfig{file=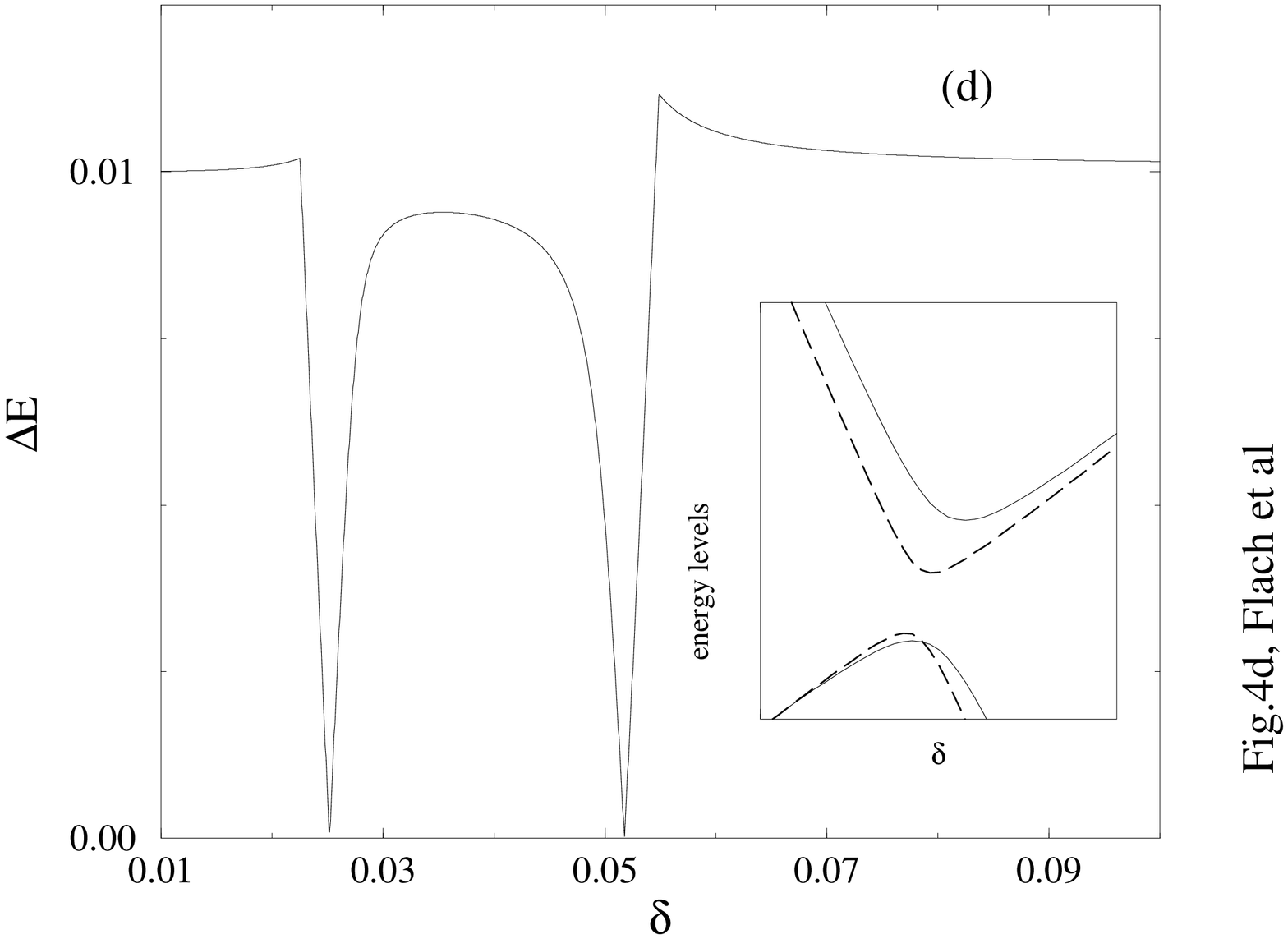,width=2.5in,angle=270}
\caption{
Level splitting variation at avoided crossings. 
\\
Inset: Variation of individual eigenvalues participating in the avoided
crossing. Solid lines - symmetric eigenstates, dashed lines - antisymmetric
eigenstates.
}
\label{fig4d}
\end{figure}

An example is shown in Fig.4(a) ($V_{s11}=V_{a11}=1$,$V_{s22}=-3$,
$V_{s12}=0.001$, $\varepsilon _{s1}=0.1$, $\varepsilon_{s2} = 0.3$,
$\varepsilon _{a1} - \varepsilon _{s1} = 0.00002$, the asymmetric level lies
below its symmetric counterpart in the tunneling pair.) where $\Delta $ is
plotted as a function of $\tilde{\delta}$. 

Crossing of a tunneling pair with an antisymmetric level is described in a
similar manner by the same matrix with interchange of indices $s$ and $a$.

\subsubsection{Pair --- pair crossing}

In order to analyze a pair --- pair crossing we should consider two tunneling
pairs with the energies $\varepsilon _{s1}$ , $\varepsilon _{a1}$, and
$\varepsilon _{s2}$, $\varepsilon _{a2}$ together with the corresponding wave
functions, $\psi _{s1}$, $\psi _{a1}$, and $\psi _{s2}$, $\psi _{a2}$. Again
we assume that these two pairs lie close to each other at $\delta = \delta
_{0}$, so that the perturbation theory can be applied, disregarding all the
other states of the trimer. Then we should consider the matrix
\begin{equation}
\left( 
\begin{array}{cccc}
\varepsilon _{1s}+V_{s11}\tilde{\delta} & V_{s12}\tilde{\delta} & 0 & 0 \\ 
&  &  &  \\ 
V_{s12}\tilde{\delta} & \varepsilon _{2s}+V_{s22} & 0 & 0 \\ 
&  &  &  \\ 
0 & 0 & \varepsilon _{a1}+V_{a11}\tilde{\delta} & V_{a12}\tilde{\delta} \\ 
&  &  &  \\ 
0 & 0 & V_{a12}\tilde{\delta} & \varepsilon _{a2}+V_{a22}\tilde{\delta}
\end{array}
\right) .  \nonumber
\end{equation}
whose diagonalization results in four energies 
\begin{eqnarray}
E_{s\{1,2\}} &=&{\frac{\varepsilon _{s1}+\varepsilon _{s2}+(V_{s11}+V_{s22})
\tilde{\delta}}{2}}\pm \sqrt{{\frac{[\varepsilon _{s1}-\varepsilon
_{s2}+(V_{s11}-V_{s22})\tilde{\delta}]^{2}}{4}}+
V_{s12}^{2}\tilde{\delta}^{2}}  \nonumber \\
&&  \label{3-17} \\
E_{a\{1,2\}} &=&{\frac{\varepsilon _{a1}+
\varepsilon_{a2}+(V_{a11}+V_{a22})\tilde{\delta}}{2}}\pm
\sqrt{{\frac{[\varepsilon _{a1}-\varepsilon
{a2}+(V_{a11}-V_{a22})\tilde{\delta}]^{2}}{4}}+
V_{a12}^{2}\tilde{\delta}^{2}}  \nonumber
\end{eqnarray}

We may now construct four possible differences between the energies of two
symmetric and two antisymmetric states which give us possible splittings. Then
choosing the difference with the lowest absolute value we define the tunneling
splitting $\Delta $ at a given value of the parameter $\delta$. The choice of
other parameters in equations (\ref{3-17}) is governed by the similar
principles as in the pair --- single level crossing. 

There are at least three different scenaria which may appear. The first two
are realized when the interaction between the two symmetric and the two
antisymmetric states is much stronger than the interaction inside each pair
which yields the individual pair splittings. If $V_{s12} \approx V_{a12}$ the
avoided crossing results in a simple resonance of the splitting (Fig. 4(b) ,
parameters
$\varepsilon _{s1}=0.1$, $\varepsilon _{s2}=0.3$,
$\varepsilon _{a1} = 0.09999$, $\varepsilon_{a2} = 0.2999$, $V_{s11} =
V_{a11}=1$, $V_{s22}=V_{a22}=-3$, $V_{s12}=V_{a12}=0.001$).

If the symmetric states are interacting stronger (or weaker) than their
antisymmetric counterparts $V_{s12} \neq V_{a12}$, two intersections inside
one of the two pairs are possible (Fig. 4(c), parameters $\varepsilon_{s1} =
0.1$, $\varepsilon _{s2} = 0.3$, $\varepsilon_{a1} = 0.09999$,
$\varepsilon_{a2} = 0.2999$, $V_{s11} = V_{a11} = 1$, $V_{s22} = V_{a22} =
-3$, $V_{s12}=0.002$, $V_{a12}=0.001$).

Finally, if the interaction between the two symmetric and the two
antisymmetric states is much weaker than the interaction inside each pair, the
pairs will intersect each other and the avoided crossing structure will appear
on a finer scale than the pair splittings. An example is shown in Fig. 4(d)
(parameters $\varepsilon_{s1} = 0.1$, $\varepsilon_{s2} = 0.3$,
$\varepsilon_{a1} = 0.09$, $\varepsilon_{a2} = 0.2$, $V_{s11} = V_{a11} = 1$,
$V_{s22} = V_{a22} = -3$, $V_{s12} = V_{a12} = 0.1$).

\section{Conclusion}

We have described and explained peculiar features of dynamical tunneling of a
strong excitation in a trimer. The tunneling amplitude may be enhanced due to
the presence of the third site with fluctuating particle number. Numerical
calculations show that this amplitude grows about eight orders of
magnitude when the coupling parameter $\delta$ still remains much smaller than
all other parameters in the system. 
An interesting feature of this process is that 
the third site serves as a donor which injects bosons into the dimer
and thus speeds up tunneling. The injected bosons are of course returned
to the third site at the end of the tunneling process.
Such a process thus needs a nonzero amount of bosons
to be deposited on the third site at the beginning of tunneling.

It is worthwhile to compare the observed enhancement of tunneling with our
consideration of a tunneling of a breather in a one-dimensional chain
\cite{fsf98}. The result of that paper was that the interaction of the
dimer with other degrees of freedom in the chain may result only in a
reduction of 
the tunneling amplitude. This finding was in agreement
with studies of disspative tunneling
\cite{lcdfgz87,w99}. 
The reason for the opposite result in the present work
is that we consider
a situation in which the third harmonic site is excited to a rather high
level and is able to provide bosons necessary to enhance the tunneling. This
is in contrast to the above references where the 
temperature was assumed to be zero
or low, meaning that the bath degrees of freedom were not excited.

Note that the discussion in Subsection II.A prompts a possibility of
a tunneling enhancement even at zero occupation of the third site. This
can be achieved in our model by choosing $g(n_3)=\omega_3 n_3$ with
$\omega_3 > B/2$. Again the expected enhancement of tunneling
is opposite to the above statement and is due to
the choice of 
such a large value of the frequency $\omega_3$
of the third site which makes the adiabatic approximation \cite{fsf98} 
invalid. It
means that the shape of the instanton calculated in \cite{fsf98} may be
significantly altered by the interaction with this high frequency mode. 

The spikes observed in the $\delta$ dependence of the tunneling amplitude
are well explained by the relevant crossings with either single levels or
with other tunneling pairs of levels. At higher values of $\delta$ when
the tunneling splitting comes close to average level spacing (1 in our
model) the whole picture becomes more complicated and one may think about an
increasing role of chaos assited tunneling \cite{btu93}. However,
discussion of this process goes beyond the framework of this paper.
\\
\\
\\
Acknowledgements.
\\
We thank D. Delande, E. Jeckelmann and Y. Zolotaryuk for
helpful discussions.

\end{document}